\documentclass[final,authoryear,3p,times,10pt]{elsarticle}

\usepackage{multirow,setspace,times,amssymb,amsmath,graphicx,color,rotating,subfigure,url}
\usepackage{lineno}
\usepackage{natbib}
\usepackage{booktabs}%
\usepackage{longtable}%

\usepackage[table]{xcolor}
\usepackage{tabularx}
\usepackage{graphicx} 
\usepackage{epstopdf}
\usepackage{mathrsfs}
\usepackage{makecell}
\usepackage[bookmarks=true,colorlinks,linkcolor=blue,anchorcolor=blue,citecolor=blue,unicode]{hyperref}
\usepackage{bookmark}

\hypersetup{CJKbookmarks=true}%
\journal{Elsevier}

\begin{document}

\begin{frontmatter}
\newpage
\title{Evolving efficiency and robustness of global oil trade networks}
\author[SB,RCE]{Wen-Jie Xie}
\author[SB]{Na Wei}
\author[SB,RCE,DM]{Wei-Xing Zhou\corref{WXZ}}
\cortext[WXZ]{Corresponding author.} 
\ead{wxzhou@ecust.edu.cn}

\address[SB]{School of Business, East China University of Science and Technology, Shanghai 200237, China}
\address[RCE]{Research Center for Econophysics, East China University of Science and Technology, Shanghai 200237, China}
\address[DM]{Department of Mathematics, East China University of Science and Technology, Shanghai 200237, China}
\begin{abstract}
As a vital strategic resource, oil has an essential influence on the world economy, diplomacy and military development. Using oil trade data to dynamically monitor and warn about international trade risks is an urgent need. Based on the UN Comtrade data from 1988 to 2017, we construct unweighted and weighted global oil trade networks (OTNs). Complex network theories have some advantages in analyzing global oil trade as a system with numerous economies and complicated relationships. This paper establishes a trading-based network model for global oil trade to study the evolving efficiency, criticality and robustness of economies and the relationships between oil trade partners. The results show that for unweighted OTNs, the efficiency of oil flows gradually increases with growing complexity of the OTNs, and the weighted efficiency indicators are more capable of highlighting the impact of major events on the OTNs. The identified critical economies and trade relationships have more important strategic significance in the real market. The simulated deliberate attacks corresponding to national bankruptcy, trade blockade, and economic sanctions have a more significant impact on the robustness than random attacks. When the economies are promoting high-quality economic development, and continuously enhancing positions in the OTN, more attention needs be paid to the identified critical economies and trade relationships. To conclude, some suggestions for application are given according to the results.
\end{abstract}

\begin{keyword}
 Global market \sep Trade network \sep Network efficiency
\\
  JEL: C1, P4, Z13
\end{keyword}

\end{frontmatter}


\section{Introduction}

As an important strategic resource, oil is often called ``black gold'' and ``blood of the modern industry''. It is not only an essential driving force for economic development but also a military and diplomatic resource. Since the 1970s, the global oil market has undergone many dramatic changes. The changes in the oil trade pattern can reflect the profound changes in the global energy system. Starting from the clues and surface phenomena, the profound underlying logic in the global oil market can be explored. Therefore, an analysis of the global oil trade network (OTN) can help us understand the current trend of political and economic development in the world. The oil trade is affected not only by the relationship between supply and demand but also by geopolitics, fluctuations in oil prices and the state of the world economy. The uneven distribution of oil resources has led to the inevitable globalization of oil trade. In such a situation, the flows of oil resources between economies constitute an important research issue concerning the global energy trading system.

Previous research on the global oil market is usually divided into two categories. Some scholars believe that the global oil market behaves as ``one great pool'', and changes in market conditions in one region can quickly affect other geographic regions \citep{Adelman-1992-Energy}. This belief is held by the majority of most oil economists. However, there is a different view that the oil market is ``regionalized'' \citep{Weiner-1991-EnJ}. In order to prove that this view is prone to mistakes, many scholars analyze the overall characteristics of the global oil market. \cite{Rodriguez-Williams-1993-ESR} test the ``one great pool'' hypothesis and the results show that the oil market is indeed a global market. \cite{Liu-Chen-Wan-2013-EM} also examine the hypothesis of ``one great pool'' by investigating trade relations between China and the four major oil markets. In contrast, \cite{Dai-Xie-Jiang-Jiang-Zhou-2016-EmpE} perform a network analysis of the crude oil spot prices all over the world and unveil regional communities. These studies prove the scientific nature of the research into global oil trade as a complex system.

As a common method for studying complex systems, complex networks have attracted the attention of experts and scholars in various fields \citep{Battiston-Farmer-Flache-Garlaschelli-Haldane-Heesterbeek-Hommes-Jaeger-May-Scheffer-2016-Science,Haldane-May-2011-Nature}. Therefore, with each economy in the oil trading system as the node and the trade relationship between the economies as the edge, the global oil trading system can be abstracted as a directed global OTN. The efficiency can be measured through topological network indexes. This study of significance in its innovative measuring of the oil flows efficiency in trade networks.

The stability and robustness of global OTN is a critical issue in the energy market. The stability of the global OTN is an essential index for judging whether the oil trade system works effectively. For the economies in the network, trade stability is an essential guarantee for economic security. Studying the stability of oil trade is of great significance for a comprehensive analysis of the state of the oil trade pattern. At the same time, the robustness indexes provide a precise method for describing and measuring the stability of oil market.

Firstly, we introduce network efficiency indexes into the global OTNs. Secondly, we identify the critical economies and critical trade relationships based on the efficiency indexes in the OTNs. The identification is the foundation for oil trade stability research. Finally, we construct network robustness indexes based on network efficiency indexes and use two types of targeted attack economy attack and trade relationship attack along with random attack and targeted attack strategies. We observe the differences in network stability under the attacks.

Analyzing the evolving efficiency and robustness of global OTNs helps understand the transmission efficiency and the stability of the network, and thereby provide policy recommendations for maintaining the oil market stability. For the government, the study can also be used as a guide for avoiding market risks. For each economy, the different criticality of each trade relationship can provide a reference for policymakers to make trade policy in the future. Since the OTNs are constantly changing and updating over time, we can observe the development trends of each economy in the OTNs through the changes of criticality indexes. The criticality indexes of trade relationships can measure the effect of relevant trade policies in each economy. For a specific trade relationship, when the trade policy is implemented, if the criticality of trade relationships increases in subsequent years, it can be considered that the trade policy promotes the contribution of trade connection to network efficiency; otherwise, the trade policy may not promote or even inhibit trade.

There are three contributions from this paper. Firstly, we introduce the network efficiency into the global OTNs and measure the effectiveness of the global oil market. Network efficiency index is commonly used in transportation networks \citep{Wandel-Sun-Cao-2012-TATS, Du-Zhou-Lordan-Wang-Zhao-Zhu-2016-TRPE}, but we use this indicator to measure the efficiency of oil trade flows in the networks and to deepen our understanding of the networks from the perspective of network efficiency. Secondly, we define the criticality indexes of economic nodes and trade relationships to examine the changes in the efficiency of the networks after the deletion of an economy node or a trade relationship. Moreover, it is beneficial for each economy to have a more comprehensive understanding of its position and critical trade relationships. Thirdly, the robustness analysis of the OTN allows us to have a deeper understanding of the rules of the collapse of the oil trading system and provides theoretical guidance for people to find better ways to prevent the collapse and maintain the oil market stability and development.

The paper is organized as follows. Section \ref{S1:LitRev} reviews the literature. Section \ref{S1:Data} describes the data source and methods. In Section \ref{S1:EmpAnal}, we construct unweighted and weighted global OTNs based on the global oil trade data to carry out an empirical analysis of the networks. Section \ref{S1:Conclude} discusses the results and concludes.

\section{Literature review}
\label{S1:LitRev}

As a complex system, the global trade network is highly interdependent, and it is difficult to understand and control \citep{Dirk-Helbing-May-2013-Nature}. Although the integrity of trade data is often a significant obstacle to the research \citep{Gleditsch-2002-JCR}, scholars are not prevented from demystifying the trading system. Early studies of trading systems based on complex network methods are often in the global scope. Staring from the perspective of the network construction method, \cite{Li-Jin-Chen-2003-PA} and \cite{Serrano-Boguna-2003-PRE} mainly focused on the unweighted network that is determined by the existence of trade relationships between economies. Later, a directional trade network was constructed based on the direction of trade flows. \cite{Fagiolo-Reyes-Schiavo-2009-PRE} and \cite{Bhattacharya-Mukherjee-Saramaki-Kaski-Manna-2008-JSM} constructed a weighted network based on the coupling of bilateral trade relationships. \cite{An-Zhong-Chen-Li-Gao-2014-Energy} built a network that can directly reflect the competition and cooperation between economies.

The topological characteristics of the global trade network have attracted many scholars \citep{Fagiolo-Reyes-Schiavo-2009-PRE, Fagiolo-2010-JEIC, Barigozzi-Fagiolo-Garlaschelli-2010-PRE, Dean-Lovely-Mora-2009-JAE, Garlaschelli-DiMatteo-Aste-Caldarelli-Loffredo-2007-EPJB}. The theoretical models proposed by \cite{Garlaschelli-Loffredo-2004-PRL} and \cite{Bhattacharya-Mukherjee-Saramaki-Kaski-Manna-2008-JSM} play a significant role in promoting the studies on the topological properties of the global trade network. Some scholars narrow their research scope to regional trade networks. \cite{Giudici-Huang-Spelta-2019-ESs} study the Asian trade network and find that the Asian trade network can be decomposed into two overlapping communities, in which different economies play different important roles. The importance of the nodes of the international trade network and the community structure derived from the calculation of the network topology indexes are important characteristics of the trade network, which are significant for studying the propagation mechanism, the evolution law, and the collectivization of trade. The studies on the importance of nodes in international trade networks mainly focus on the analysis of node centrality indexes \citep{Dablander-Hinne-2019-SR, Richmond-2015-SSEP}, such as node degree, betweenness, and closeness centrality. Some scholars have also proposed a new index to measure the importance of network nodes, such as \cite{Battiston-Puliga-Kaushik-Tasca-Caldarelli-2012-SR}, who introduce DebtRank as a system influence index for nodes.

Scholars mostly adopt the method of studying global trade networks in the study of OTNs. \cite{Zhang-Lan-Xing-2018-IOP} apply the complex network method to analyze the average transaction volume, network connectivity, network division and major community economies of the networks. The result proves that the connectivity of the petroleum product network is higher than that of the crude oil network. The result also proves that geographical factors are increasingly more obvious in the crude oil trade pattern, and the choice of trading partners of various oil trade economies is affected by factors such as geographical resistance. \cite{Kitamura-Managi-2017-AEn} believe that geopolitical resistance has restricted economies from choosing similar oil exporters. The description of the flow of oil trade with the gravity equation displays that bilateral trade is proportional to the gross product of both parties and inversely proportional to the distance between them. \cite{Kharrazi-Fath-2016-EP} use the PMI (point-wise mutual information) index to study the dependent relationships in the OTNs. They find that the PMI index can reflect the interdependence between economies over time. \cite{An-Zhong-Chen-Li-Gao-2014-Energy} find that the international crude oil trade is evolving into a stable, orderly and integrated system, and different types of events have different effects on importing and exporting countries. Based on the dependence of global oil trade, \cite{An-Wang-Qu-Zhang-2018-Energy} study the changes in the international dependency network after the unexpected oil prices drop in 2014, revealing the impact of the global oil market after the shock of oil prices not only at the level of various economies but also at the system level. \cite{Gao-Sun-Shen-2015-AppliedEnergy} discuss several important indexes such as degree distribution, community stability and importance of major countries, and analyze the scale-free and pattern evolution characteristics of the fossil fuel trade network. Against the backdrop of intensifying global competition, \cite{Yu-Jessie-Sharmistha-2015-Energy} analyze the geography and evolution of global oil flows. Cooperation and competition are also hot topics in global oil trade research. \cite{Zhang-Fu-Pu-2019-CPr} provide suggestions for cooperation and sustainable development of the Chinese ``Belt and Road'' oil trade. \cite{Zhang-Ji-Fan-2014-EP} analyze the evolution and transmission of competition modes among oil-importing countries, taking into account the strength of competition indicators.

\cite{Du-Wang-Dong-Tian-Liu-Wang-Fang-2017-AEn} assess the relative importance of each economy in the trade network by constructing top-level import and export trade networks. \cite{Zhong-An-Shen-Fang-Gao-Dong-2017-EP} analyze the international fossil fuel trade network by examining top-level relations, centrality, intermediary capabilities of each country, and the role of each country in trade groups. The bridge function of the central node in the process of trade network formation is also proved \citep{Ji-Zhang-Fan-2014-ECM}.

As is often the case with global networks such as financial networks, global OTNs face systemic risks. Systematic risk is difficult to define and measure. The interrelationships between the various entities in the system make the risk contagious. It is also believed that traditional economic theories cannot well explain and predict the financial system collapse and its continuing impact on the global economy \citep{Battiston-Farmer-Flache-Garlaschelli-Haldane-Heesterbeek-Hommes-Jaeger-May-Scheffer-2016-Science}. Due to the complexity and instability of global network systems, the factors affecting system risk include many aspects \citep{Haldane-May-2011-Nature, Munnix-Shimada-Schafer-Leyvraz-Seligman-Guhr-Stanley-2012-SR, Vodenska-Aoyama-Fujiwara-Iyetomi-Arai-2016-PLoS1, Acemoglu-Ozdaglar-TahbazSalehi-2015-AER, Acharya-Pedersen-Philippon-Richardson-2017-RFS}. For example, based on the oil export trade data of 34 major oil-exporting economies, \cite{Du-Dong-Wang-Zhao-Zhang-Vilela-Stanley-2019-Energy} construct an oil import network with time evolution characteristics and analyze the structural changes of the correlation network. It is found that abrupt percolation transition would lead to spikes in systemic risk. Identifying the abrupt percolation transition can alert systemic risks 3-11 months in advance.

Systemic risk arises from the cascade effect of the failure of economies or trade relationships in the networks. \cite{Sun-Gao-Zhong-Liu-2017-PA} study the impact of economies and trade relationships on the stability of the global OTNs. \cite{Zhong-An-Gao-Sun-2014-PA} analyze the evolution characteristics and stability of the topological indexes of the unweighted and weighted OTNs and find that different causes of instability have different characteristics. \cite{Ji-Zhang-Fan-2014-ECM} analyze the stability of oil trade in the case of trade interruption at both global and national levels. They conclude that the OTN can be characterized by ``robust but fragile''.

The network robustness describes the ability of a network to maintain certain structural integrity and functions in the event of random failure of nodes and edges or deliberate attacks. It is an essential dynamic characteristic of the network system. Therefore, the more robust a network is, the more stable it is.  In this sense, what enhances the ability to resist systemic risks is actually to enhance the robustness of the network system. In order to measure the robustness of the network system, \cite{Latora-Marchiori-2001-PRL} propose a network efficiency index to measure the efficiency of information exchange and accurate quantitative analysis of information flow. The efficiency index lays a good foundation for the research on the robustness and vulnerability of transportation networks \citep{Criado-Hernandez-Bermejo-Romance-2007-IJBC, Dall'Asta-Barrat-Barthelemy-Vespignani-2006-JSM, Wandel-Sun-Cao-2012-TATS, Du-Zhou-Lordan-Wang-Zhao-Zhu-2016-TRPE}. Based on the efficiency of the network, \cite{Crucitti-Latora-Marchiori-Rapisarda-2002-PA} analyze the impact of the attacks on the global efficiency and local efficiency of the network. Network efficiency also becomes an entry point for scholars to study the cascade effect. \cite{Crucitti-Latora-Marchiori-2004-PRE} propose a simple dynamic redistribution cascade fault model based on network flow. The study proves that when one node is one of the most heavily loaded, the collapse of a single node is enough to crash the efficiency of the entire system. Based on the concept of network efficiency and robustness, \cite{Latora-Marchiori-2007-NJP} propose a new index to measure the centrality and prove that the centrality index based on the efficiency index is not only suitable for weighted and unweighted networks, but also networks between groups and individuals.

The network efficiency index can measure the efficiency of information flows in the information networks and the transportation efficiency in the transportation networks. Then can the efficiency of trade flows in the OTN be measured by the network efficiency index? We introduce the network efficiency index into the global OTNs. The network efficiency index is helpful for understanding the evolution of oil flows efficiency over time. According to the network efficiency index, the robustness of the networks can be studied, and networks with strong robustness are more stable. In robustness analysis, the critical issue is to identify critical nodes or edges, all of which have a critical impact on the robustness of networks. We can calculate corresponding measures, such as strengthening the supervision of these critical nodes or constructing alternative connections. Scholars have achieved inspiring results in the identification of important nodes and essential network edges \citep{Chen-Lam-Sumalee-Li-Li-2012-CEPR} based on network efficiency index\citep{Zhou-Wang-Hang-2019-TRPE}.


\section{Data and methodology}
\label{S1:Data}

\subsection{Data}

\subsubsection{Global oil trade data from UN Comtrade}

The global oil trade data of economies and regions in this study comes from UN Comtrade, and the data code is HS270900. Since the data comes from the official data reported by the trading parties, there are inconsistencies. We select more complete oil trade import data for trade relations. We extract the required data from the raw data, including time, export economy, import economy and trade volume. The trade relationships and trade volumes between the various economies are extracted annually.

The top economies in terms of oil imports and exports occupy a vital position in the OTNs. According to the data, we make statistics on the top 10 economies in the total trade volume of oil exports in the trade network from 1988 to 2017.

\begin{figure}[!t]
\centering
\includegraphics[width=0.95\linewidth]{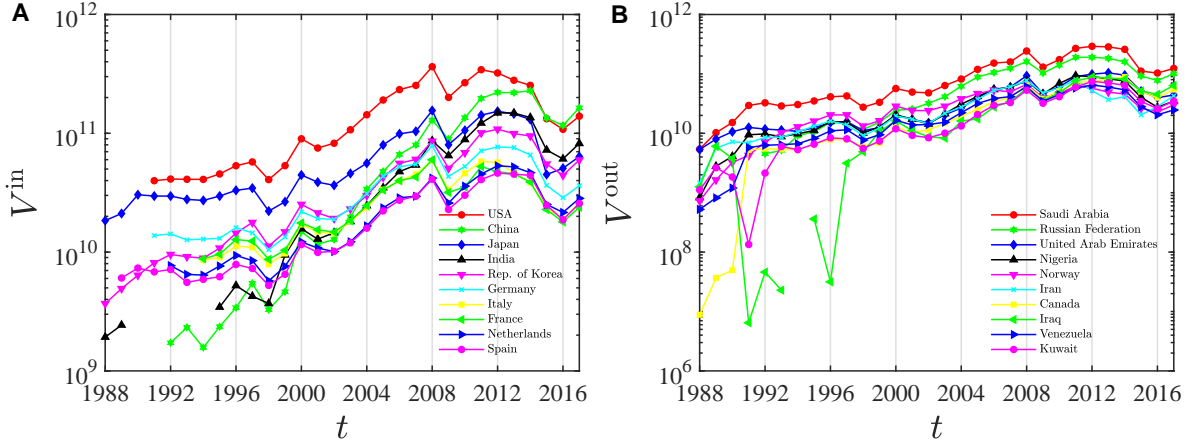}
  \caption{
  The evolution of export and import trade volumes of the top 10 economies from 1988 to 2017. We calculate the total volume of import and export by adding up the annual volume of 258 economies. Then, we take the top 10 economies in terms of total import and export volumes. The vertical axis is the logarithm of the import and export volumes. The evolution of oil trade for the top 10 economies in total oil import $V^{\rm{in}}$ and export $V^{\rm{out}}$ trade volumes from 1988 to 2017 are shown in (A) and (B). Each solid line represents the evolution curve of an economy's trade volume.}
    \label{Fig:Import:Export:year}
\end{figure}

It can be seen from Fig.~\ref{Fig:Import:Export:year}, the total volumes of import and export trade of various economies have increased by the year. In Fig.~\ref{Fig:Import:Export:year}~(A), the economy with the largest import volume is the USA, followed by China and Japan. The total oil trade of each economy has shown a downward trend since 2011, suggesting that the trend of oil trade and network structure is changing as the global oil supply and demand structure is adjusted, and the development of the refining industry is accelerated. The total volume of China's oil imports rose the most from 1992 to 2017. Prior to 2014, the economy which ranked first in terms of total import is the USA. However, China surpasses the USA to become the first economy from 2015 to 2017. The ranks of other economies are relatively stable, and the trade volumes have a similar trend over time. The economy with the highest export volume in Fig.~\ref{Fig:Import:Export:year}~(B) is Saudi Arabia, followed by Russia, the United Arab Emirates, and so on. The total oil export volumes of Saudi Arabia and Russia are far ahead of other economies. Among the top 10 economies, more than half of the economies play a unique role as exporters in the OTNs, such as Saudi Arabia and Iraq. Some economies such as Russia and Nigeria also import oil from a small number of economies. The establishment of trade relationships between economies is not only related to the need of the economy to maintain a certain amount of oil reserves but also influenced by political factors and mutual restraint between economies.

\subsubsection{Construction of global OTNs}

The global OTN is a complex system. As long as there is a trade relationship between the economies, there is an edge in the network. Because of globalization, there are many complicated trade relationships in global OTNs. We define two types of directed networks. The first type is the unweighted OTN $A_{t}$ in year $t$, which is represented by the adjacency matrix $A$. The row and column of the matrix elements represent the export and import relationship, respectively. The matrix element $a_{ij}=1$ indicates that oil from economy $i$ is sold to economy $j$. The second type is the weighted OTN $W_{t}$ in year $t$, which is represented by the matrix $W$. The matrix element $w_{ij}=v_{ij}$ indicates that oil from economy $i$ is sold to the economy $j$, and the trade volume of oil is $v_{ij}$. The total export volume of the economy $i$ in the weighted OTN is $V_{i}^{\rm{out}}$,
\begin{equation}\label{Eq:value:out}
V_{i}^{\rm{out}}=\sum_{j=1}^{N}v_{ij}
\end{equation}
The total import volume of the economy $i$ in the weighted OTN is $V_{i}^{\rm{in}}$,
\begin{equation}\label{Eq:value:in}
V_{i}^{\rm{in}}=\sum_{j=1}^{N}v_{ji}
\end{equation}
where $N$ is the number of economies in the trade network, the total volume of the import and export of an economy in the OTN is the sum of the trade volumes with other economies. From 1988 to 2017, the total trade volume of the entire oil trading network is represented by $V$ each year.

\begin{figure}[!t]
\centering
\includegraphics[width=0.47\linewidth]{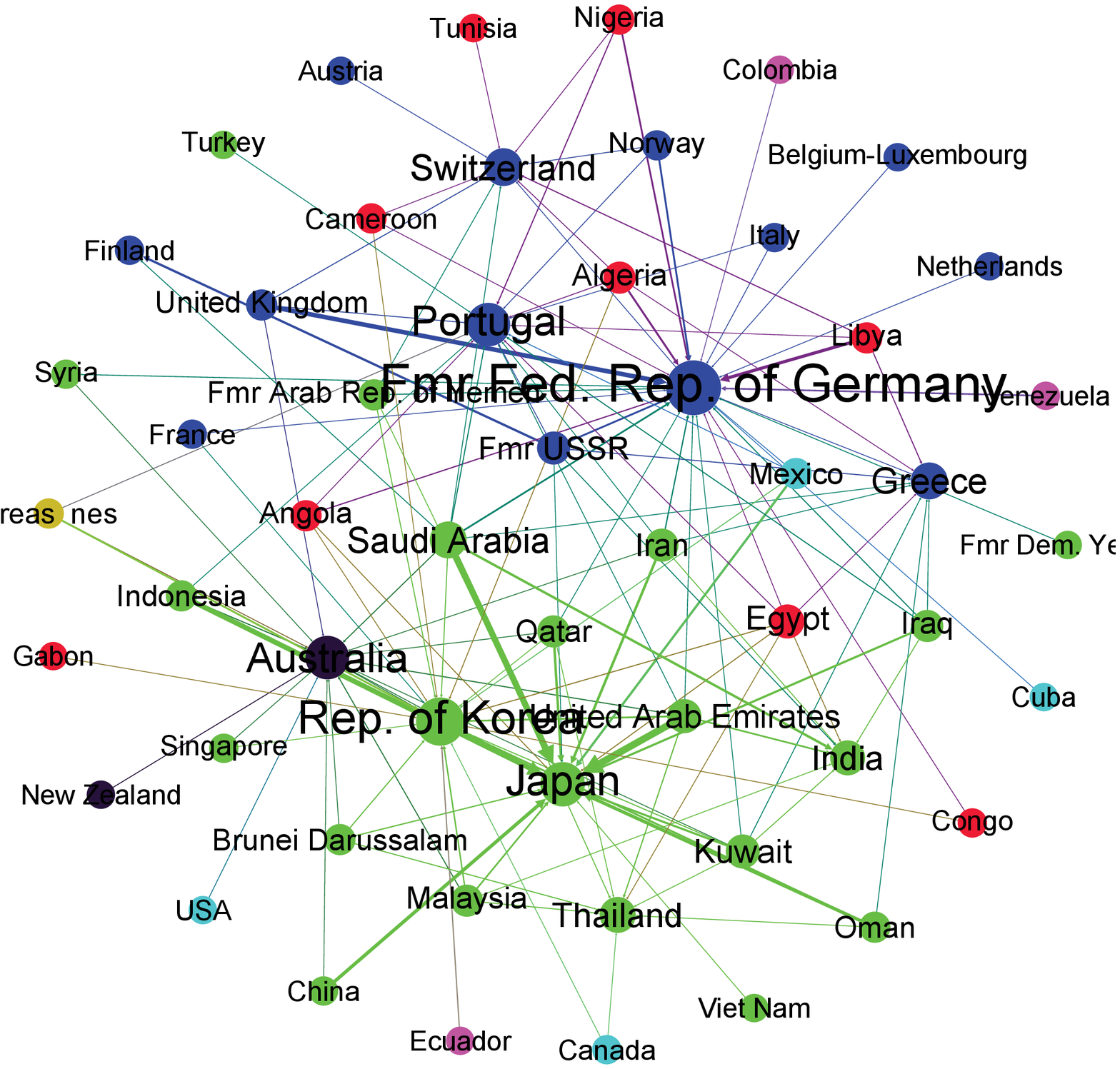}\hspace{3mm}
\includegraphics[width=0.47\linewidth]{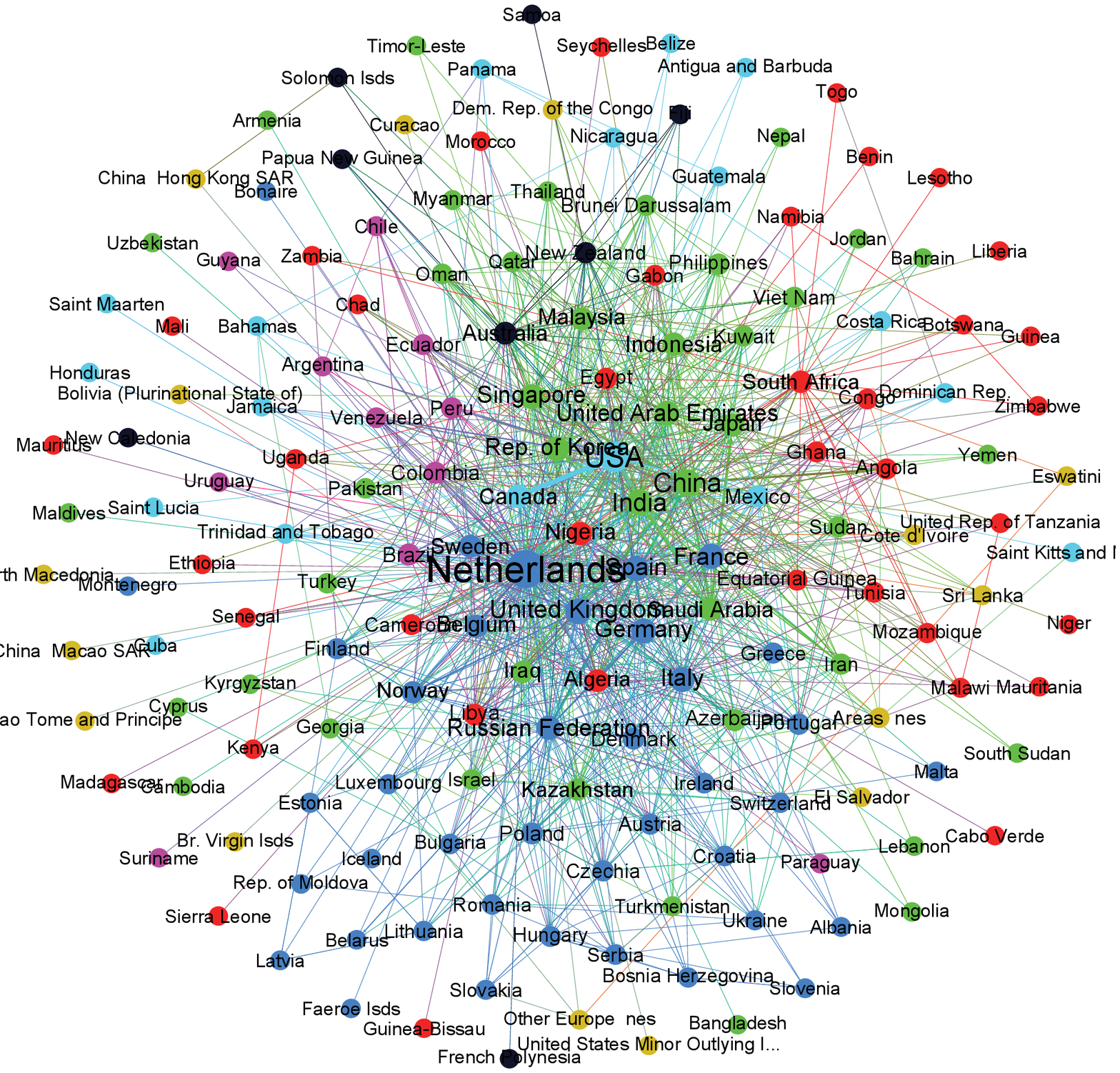}
\caption{
The oil trade networks of 1988 (A) and 2017 (B). The size of the nodes and labels in the figure is determined by the degree of the nodes. The larger the degree, the larger the node. Nodes of different colors represent six different continents.} \label{Fig:1988:2017:Network}
\end{figure}

We construct 30 OTNs for the period from 1988 to 2017 and define $N_e$ as the total number of trade relationships. As can be seen from Fig.~\ref{Fig:1988:2017:Network}, the OTN in 1988 was quite different from 2017. The picture on the left shows the OTN in 1988. Earlier in 1988, there is less data and fewer nodes in the network. The picture on the right is the network in 2017, revealing that in comparison with 1988, nodes increase more rapidly, the number of trade relationships is more enormous, and the trade relationships are more complicated.

\subsection{Methodology}

\subsubsection{Efficiency of global OTNs}

There are many indexes to evaluate and compare the efficiency of the entire network from the network structure, such as the maximum principal component of the network and the average shortest path length. In order to overcome the limitations of these indexes, \cite{Latora-Marchiori-2001-PRL} first propose network efficiency, which is used to measure the efficiency of information exchange in the network and to perform accurate quantitative analysis of information flows. In the OTN, the efficiency index is used to measure the efficiency of trade flows. The efficiency of the unweighted trade network $A_t$ is $E^{A}$, which can be obtained by the following formula
\begin{equation}\label{Eq:unweight:efficiency}
E^{A}=\frac1{N(N-1)}\sum_{i\neq j\in A_t}\frac1{d_{ij}},
\end{equation}
where $d_{ij}$ is the shortest path between economy $i$ and $j$, and $N$ is the total number of nodes in the network $A_t$. The unweighted network efficiency is defined as the average of the reciprocal of the shortest path length. The reciprocal of the shortest path length can be seen as the ``efficiency'' of this path. Eq.~(\ref{Eq:unweight:efficiency}) does not consider weight when calculating network efficiency. We can generalize this situation to the weighted network $W_t$, whose efficiency $E^{W}$ can be defined as
\begin{equation}\label{Eq:weight:efficiency}
E^{W}=\frac1{N(N-1)}\sum_{i\neq j\in W_t}e_{ij},
\end{equation}
where $e_{ij}$ is the path efficiency between economy $i$ and $j$. In an unweighted network $e_{ij}=\frac1{d_{ij}}$, the shorter the path length, the higher the efficiency. In the weighted network, we need consider the weights of the edges. The weight is the volume of the oil trade between the two economies. So the path efficiency between the two economies $i$ and $j$ is
$e_{ij}=\frac1{\sum_{l\in L_{ij}}{\frac1{v_{l}}}}$,
where $L_{ij}$ is the set of edges for the shortest path between the economies $i$ and $j$ in the weighted network $W_t$, $l$ is the element in $L_{ij}$, and $v_{l}$ is the weight of the edge $l$. Then we can define the efficiency of the weighted network $W_t$ as
\begin{equation}\label{Eq:finial:weight:efficiency}
E^{W}=\frac1{N(N-1)}\sum_{i\neq j\in W_t}e_{ij}=\frac1{N(N-1)}\sum_{i\neq j\in W_t}\frac1{\sum_{l\in L_{ij}}{\displaystyle\frac1{v_l}}}.
\end{equation}

From Eq.~(\ref{Eq:finial:weight:efficiency}), it can be seen that with the increase of the trade volume $v$, the network efficiency increases. In order to analyze the effect of network structure on effectiveness, we redefine a new network efficiency index by standardizing all the weights,
\begin{equation}\label{Eq:weight:norm}
\bar{v}_l=\frac{v_l}{\langle v \rangle},
\end{equation}
$\langle v \rangle$ is the average volume of the trade relationships. By substituting the parameter $\langle v \rangle$ into the $v_l$ in Eq.~\ref{Eq:finial:weight:efficiency}, we can define the index $E^{\bar{W}}$ is
\begin{equation}\label{Eq:weight:norm:efficiency}
E^{\bar{W}}=\frac{E^{W}}{\langle v \rangle}.
\end{equation}

Therefore, the efficiency of different years is comparable. $E^{\bar{W}}$ can reflect the influence of network structure on network effectiveness.

\subsubsection{Criticality of oil trade economies}

The economies in the OTNs are heterogeneous and on different positions. If critical economies in the network are removed, the efficiency of the networks will be greatly reduced. Therefore, according to network efficiency indicators, important economies in the network can be identified. We define $G_{i}$ as the network after removing the economy $i$ and all its relationships from the original OTN $A_t$ and $W_t$. For $A_t$, we define the criticality of the economy $i$ as $C_i^{A}$, and the weighted network $W_t$ is $ C_i^{W}$, and then we have
\begin{equation}\label{Eq:node:criticality}
C_i^{A}=1-\frac{E^{A}(G_i)}{E^{A}} ~~~ and ~~~ C_i^{W}=1-\frac{E^{W}(G_i)}{E^{W}}.
\end{equation}
where $E^{A}(G_i)$ and $E^{W}(G_i)$ are the network efficiency of the unweighted and weighted OTN after deleting the economy $i$ and all its relationships.

According to Eq.~(\ref{Eq:node:criticality}), we can calculate the criticality ranking of each economy in network $A_t$ and $W_t$. The rankings represent the strength of each economy's influence on efficiency.

\subsubsection{Criticality of oil trade relationships}

With the advancement of economic globalization, the total number of economies in the global OTNs has shown a steady upward trend. The rapid changes in global oil trade often reflect the changing relationships among economies, and it is of practical significance to measure which trade relationships are more critical to the efficiency of trade networks. According to network efficiency, we can identify critical economies in the network. Similarly, if the critical trade relationships in the network are removed, the efficiency of the trade network will be greatly reduced. The removal of trade relationship corresponds to economic sanctions in the real world. We define $G_{ij}$ as the network after removing the relationship $i$ to $j$ from the original OTN $A_t$ or $W_t$. The criticality of the relationship from $i$ to $j$ is
\begin{equation}\label{Eq:edge:criticality}
C_{ij}^{A}=1-\frac{E^{A}(G_{ij})}{E^{A}} ~~~ and ~~~ C_{ij}^{W}=1-\frac{E^{W}(G_{ij})}{E^{W}}.
\end{equation}
where $E^{A}(G_{ij})$ and $E^{W}(G_{ij})$ are the network efficiency of the unweighted and weighted OTNs after deleting the relationship between $i$ and $j$. According to Eq.~(\ref{Eq:edge:criticality}), we can calculate the criticality rankings of all trade relationships in the global OTNs $A_t$ and $W_t$. Criticality rankings represent the strength of the influence of trade relationships on the efficiency of the global OTNs.

\subsubsection{Robustness of the global OTNs}

The OTNs are affected by many factors, such as economic and financial events and wars. In most researches on network robustness, network attacks are generally classified into two types: node attack and edge attack. We consider the robustness of unweighted and weighted OTNs under the two attacks. First, we define the robustness index based on the network efficiency
\begin{equation}\label{Eq:robustness:node:edge:attack}
R_{\alpha}^{\beta}(p)=\frac{E^{\beta}(G_{\alpha}^{\beta}(p))}{E^{\beta}(G^{\beta})}.
\end{equation}

$G^{\beta}$ is the network $A_t$ or $W_t$ before the attack, and $G_{\alpha}^{\beta}(p)$ is the network $A_t$ or $W_t$ after being subjected to node attack and edge attack under each strategy. $p$ is the proportion of deleted economies or relationships. It should be noted that when an economy is attacked, we also remove the relationships associated with the economy from the network. $E(G(p))$ is the efficiency of the OTN after economies and relationships being removed with the proportion of p. According to Eq.~(\ref{Eq:robustness:node:edge:attack}), the robustness of network $A_t$ and $W_t$ after being subjected to various attacks can be calculated. Under different network types, different attacks can be defined by different values of $\alpha$ and $\beta$.

In Eq.~(\ref{Eq:robustness:node:edge:attack}), $\beta$ defines whether the network robustness is of the network $A_t$ or of $W_t$. In addition, it defines whether the attack on the network is node attack or edge attack. In order to define these four attributes, we use $R^{A}$ and $R^{W}$ to represent node attack on $A_t$ and $W_t$. $R^{a}$ and $R^{w}$ represent edge attack on $A_t$ and $W_t$.

$\alpha$ in Eq.~(\ref{Eq:robustness:node:edge:attack}) defines the 5 specific strategies adopted for the attack. $\alpha$ can be $\rm{random}$, $\rm{criticality}$, $\rm{in}$, $\rm{out}$, and $\rm{value}$. $R_{\rm{random}}$ means randomly selecting economies or relationships in the network to attack; $R_{\rm{criticality}}$ means attacking the network in order of the criticality of the economy or relationships in the network from high to low, which means the more critical economies and trade relationships will be selected and attacked firstly. $R_{\rm{in}}$ and $R_{\rm{out}}$ only indicate the attack strategy when the economy is attacked, with $R_{\rm{in}}$ and $R_{\rm{out}}$ representing attacks on economy from high to low based on the rankings of total import and export volumes. $R_{\rm{value}}$ means attack on the relationships of the weighted network with the weight from high to low.

For the random attack strategy, we need to give the calculation formula
\begin{equation}\label{Eq:random:robustness}
R_{\rm{r}}^{\beta}(p)=\frac1{\frac{N!}{N_p!(N-N_p)!}}\sum_{G_{\rm{random}}^{\beta}(p)}\frac{E^{\beta}(G_{\rm{random}}^{\beta}(p))}{E^{\beta}(G^{\beta})},
\end{equation}
where $\frac{N!}{N_p!(N-N_p)!}$ is the number of possible combinations when $N_{p}$ economies or relationships are randomly selected from the OTN $A_t$ or $W_t$. $\frac{N!}{N_p!(N-N_p)!}$ may be large, and we use Monte Carlo method to estimate $R_{\rm{random}}^{\beta}(p)$ in all cases.

By calculating the robustness of the OTNs under various attack strategies, we can calculate the stability in various situations.

\section{Empirical analysis}
\label{S1:EmpAnal}

\subsection{Evolving efficiency of global OTNs}

\begin{figure}[!t]
\centering
\includegraphics[width=0.95\linewidth]{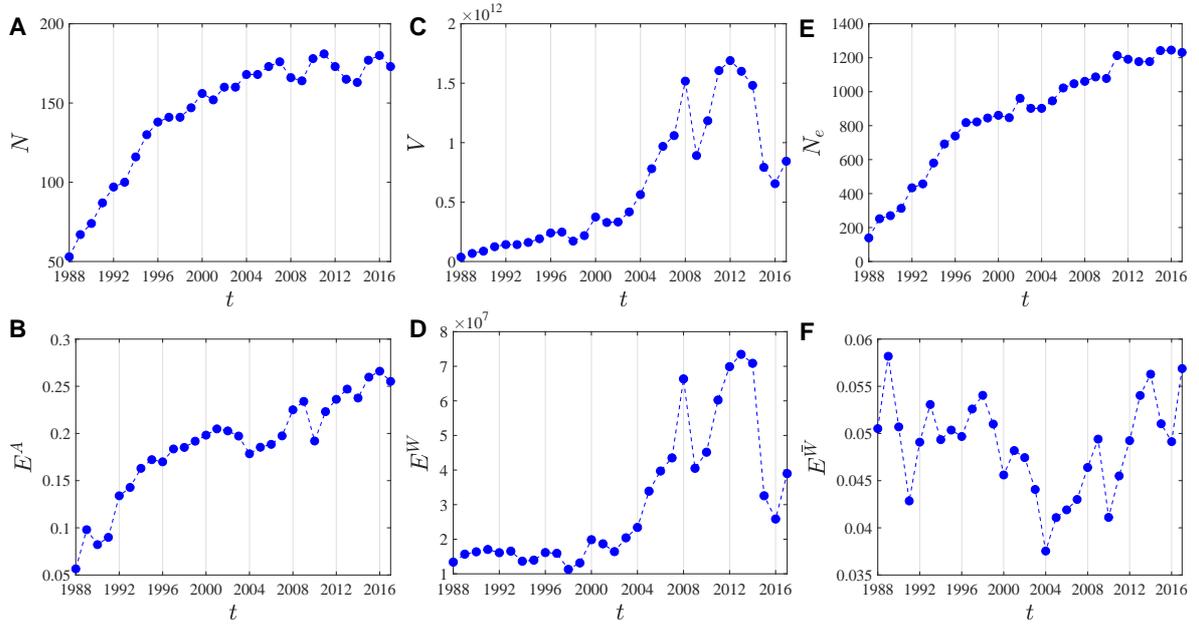}
  \caption{
  Evolution of the number of economies, relationships and network efficiency. From 1988 to 2017, (A) is the evolution of the number of economies $N$ in the global OTNs. (B) is the evolution of the efficiency $E^{A}$ of $A_t$ calculated according to Eq.~(\ref{Eq:unweight:efficiency}). (C) shows the evolution trend of the total oil trade volume of the OTNs from 1988 to 2017. (D) is the efficiency evolution of $W_t$ , calculated according to Eq.~(\ref{Eq:weight:efficiency}). (E) shows the evolution of the number of relationships in OTNs. $N_e$ is the number of relationships. (F) is the evolution of the efficiency after the weights are standardized according to Eq.~(\ref{Eq:weight:norm:efficiency}).}
    \label{Fig:oil:efficiency:nodes:edges:value}
\end{figure}

The weighted efficiency we proposed compares the distribution of weights between the weighted and unweighted networks. Some basic attributes and network efficiency of the two types of OTNs are shown in Fig.~\ref{Fig:oil:efficiency:nodes:edges:value}. It can be seen from Fig.~\ref{Fig:oil:efficiency:nodes:edges:value}~(B), the efficiency of the network $A_t$ is generally on the rise over time, and decline significantly in specific years, such as 1990, 2010 and 2014. Fig.~\ref{Fig:oil:efficiency:nodes:edges:value}~(A,E) and Fig.~\ref{Fig:oil:efficiency:nodes:edges:value}~(B) are similar in terms of trends, which can also explain that with the increase of economies and trade relationships in the network, the efficiency of network trade flows is also increasing.

For the weighted network $W_t$, we sum up the trade volumes of all economies in each year and show the result in Fig.~\ref{Fig:oil:efficiency:nodes:edges:value}~(C). The total volumes fall sharply in 2009, which may be due to the impact of the global economic downturn after the global financial crisis in 2008. The total volumes gradually rose after 2009 and reached its highest point in 2012, and it fell back to the lowest point in 2016. From the evolution of total volumes, it can be seen that the OTNs are more unstable in recent years, which confirms the complexity of global oil trade relationships and the advance of globalization.

Fig.~\ref{Fig:oil:efficiency:nodes:edges:value}~(D) is the weighted network efficiency calculated by Eq.~(\ref{Eq:weight:efficiency}). Considering the weight of the OTNs, the network efficiency fluctuates greatly in some years. The weights of networks determine $ E^{W}$. Compared with Fig.~\ref{Fig:oil:efficiency:nodes:edges:value}~(B), Fig.~\ref{Fig:oil:efficiency:nodes:edges:value}~(D) shows greater fluctuations and is similar to the trend of the evolution of the total volumes of networks in Fig.~\ref{Fig:oil:efficiency:nodes:edges:value}~(C). For example, the efficiency of the OTNs after 2009 and 2016 has a significant decline, which indicates that major events such as the financial crisis have a significant impact on the OTNs.

Finally, in Fig.~\ref{Fig:oil:efficiency:nodes:edges:value}~(F), we calculate the network efficiency after the weights are standardized according to Eq.~(\ref{Eq:weight:norm:efficiency}), rendering the efficiency more comparable. Different from Fig.~\ref{Fig:oil:efficiency:nodes:edges:value}~(B) and Fig.~\ref{Fig:oil:efficiency:nodes:edges:value}~(D), the efficiency in Fig.~\ref{Fig:oil:efficiency:nodes:edges:value}~(F) fluctuates considerably. There is no distinct tendency, which also reflects the differences in network efficiency in different years. No matter which kind of network efficiency is applied, the OTN efficiency is described from different perspectives.

\subsection{Evolving criticality of economies in global OTNs}

\subsubsection{Evolving criticality of each economy}

\begin{figure}[!t]
\centering
\includegraphics[width=0.95\linewidth]{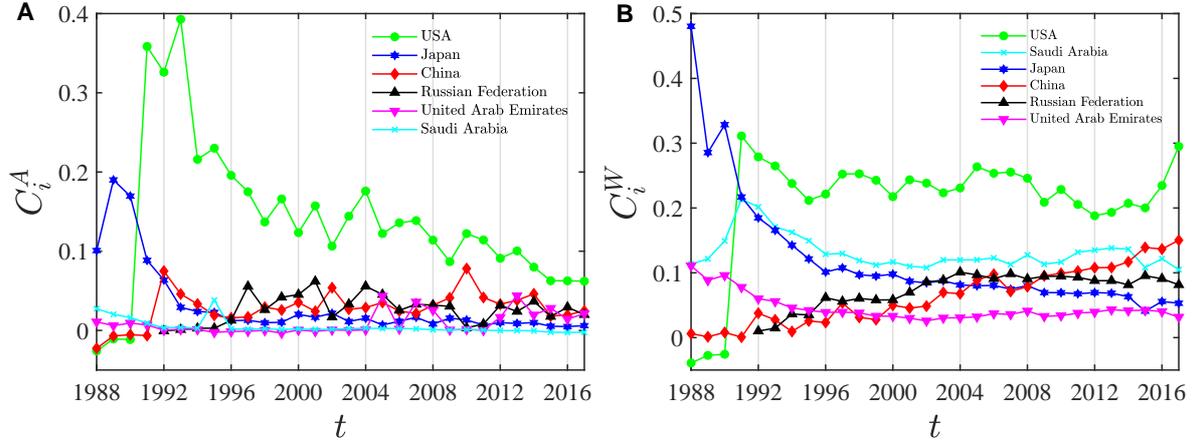}
  \caption{
  Criticality evolution of each economy in unweighted network $A_t$ and weighted network $W_t$. (A) and (B) are the economic criticality calculated according to Eq.~(\ref{Eq:node:criticality}), and they show the criticality evolution of economy $C_i^{A}$ and $C_i^{W}$. The lines represent the evolution of the criticality of the top three economies in terms of total import and export volumes, ranked in descending order according to the criticality of the six economies for 30 years.}
    \label{Fig:oil:criticality:evolution}
\end{figure}

Based on the criticality indexes of the economies in the global OTNs, Fig.~\ref{Fig:oil:criticality:evolution} shows the criticality of the economy in network $A_t$ and $W_t$. Since the number of economies involved in the trade networks is large, we select six economies whose total volumes of import and export are ranked the top 3. From the lines in the two figures, we can see that the criticality of the economies is quite different. Seen from Fig.~\ref{Fig:oil:criticality:evolution}~(A) and Fig.~\ref{Fig:oil:criticality:evolution}~(B), the USA is in a leading position. According to $C_i^{A}$, the criticality of the USA shows a downward trend, which may be due to the increase in the number of economies and trade relationships and the complexity of the networks. The number of economies with more trade relationships in unweighted networks has increased, and the structure criticality of the USA has declined. Considering the network structure with weight, the criticality of the USA is relatively stable, with small fluctuations. Such fluctuations may be affected by events such as the financial crisis and the European debt crisis. In addition, the top 3 economies have more fluctuations when they are viewed according to the unweighted criticality index than when they are viewed according to the weighted criticality, such as China and Russia. From Fig.~\ref{Fig:oil:criticality:evolution}~(B), it can be seen that with the exception of the USA and Japan, the trend of the criticality of each economy is stable. Japan's criticality surpassed the USA before 1990 and then declined until it reached a plateau. Because the amount of data in previous years is small and may be incomplete, the indexes are relatively volatile. The weighted criticality of China and Russia rise steadily, reflecting the development trend of the two economies in the networks and their growing criticality in the international oil market.

\begin{figure}[!t]
\centering
\includegraphics[width=0.95\linewidth]{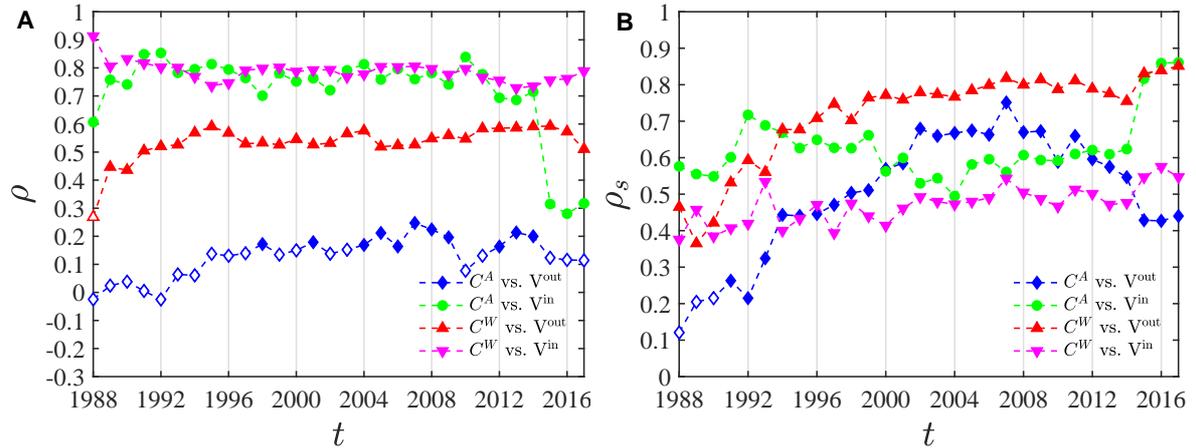}
  \caption{
  The correlation between the criticality of various economies and the volumes of import and export trades and the evolution of their significance levels. (A) and (B) are the evolution trend of Pearson $\rho$ and Spearman correlation coefficient $\rho_s$ of the criticality and the volumes from 1988 to 2017.
  }
    \label{Fig:oil:criticality:value:corr}
\end{figure}

Considering the similarity between weighted network efficiency and network weights reflected in Fig.~\ref{Fig:oil:efficiency:nodes:edges:value}~(C) and Fig.~\ref{Fig:oil:efficiency:nodes:edges:value}~(D), the criticality is defined based on the network efficiency. We make a correlation analysis between the economic criticality and the trade import and export volumes of the economies. The result is shown in Fig.~\ref{Fig:oil:criticality:value:corr}. In Fig.~\ref{Fig:oil:criticality:value:corr}~(A), it can be seen that the correlation coefficients fluctuate smoothly in the time interval. There is a significant positive correlation between the import volumes and the criticality. The export trade volumes have a smaller significant positive correlation with the criticality of the economy in $W_t$. There is no significant correlation between the volume of export trade and the criticality of economies in $A_t$. In Fig.~\ref{Fig:oil:criticality:value:corr}~(B), there is a high degree of positive correlation between the export volumes and the criticality in $W_t$, and the correlation of import volumes and the criticality of the economies in $A_t$ is significant. There is no significant correlation between the export volumes and the criticality of the economies in $A_t$ from 1988 to 1990. The correlation coefficient between the weight criticality and import volumes of the economies is between 0.4 to 0.6.

There is a correlation between the criticality of the economies and the volumes of import and export because of the consideration of the edge weights in the calculation of network efficiency indexes. Despite only considering the relationships in the network $A_t$, economies with more trade relationships are more likely to have larger trade volumes. Then there is a specific relationship between economic criticality in $A_t$ and trade volumes. However, the weighted criticality index measures the impact of each economy on network efficiency from the perspective of network structure and trade volumes. Compared with measuring the status of each economy in the OTNs relying solely on trade volumes or the previously proposed centrality indexes, the criticality index of economy provides a new perspective.

According to Eq.~(\ref{Eq:node:criticality}), we calculate the unweighted and weighted criticality. The top 10 economies in each year are shown in the tables below. We examine the evolution of the importance of each economy from the perspective of the criticality index of each economy in networks. From Table~\ref{tab:node:criticalityRank10} and Table \ref{tab:node:weight:criticalityRank10}, we obtain the following findings.

  The rankings of economic criticality calculated in unweighted and weighted OTNs are quite different. For $A_t$, critical economies may have a higher criticality because they have a higher degree. Taking into account the weight of the networks, economies with more trade volumes will have higher criticality. The weighted criticality index can integrate the structural characteristics of the network and the strength of the trade relationships.

  After 1990, Japan was replaced by the USA as the most critical economy in the OTN. The USA is identified as the most critical economy in most years, which is in accordance with its most critical developed country identity. In recent years, the international oil situation has become complicated, and the USA has gradually moved towards energy self-sufficiency from the largest oil importer in the past. \cite{An-Wang-Qu-Zhang-2018-Energy} find that due to complex geopolitical relationships and ongoing regional conflicts, the USA has significantly reduced its dependence on traditional oil-producing regions in the Middle East and Africa. This may be a significant reason why the USA has changed its criticality ranking in the unweighted trade networks in recent years.

  Comparing Table~\ref{tab:node:criticalityRank10} with Table~\ref{tab:node:weight:criticalityRank10}, we can conclude that from 2015 to 2017, the Netherlands becomes the most critical country in the network $A_t$, while the USA is still the most critical economy in the network $W_t$. This result may be explained by that the economic criticality index in the network $A_t$ focuses more on the network structural characteristics. In the past three years, the Netherlands had more trade relationships and became the largest economy in its community \citep{An-Wang-Qu-Zhang-2018-Energy}. Moreover, its trade flows efficiency is more critical than the USA. In the network $W_t$, the USA still occupies the most critical position in the network because of its vast transaction volume and complex trade relationships. The Netherlands is not among the top economies in network $W_t$.

\begin{table*}[!t]
  \centering
  \caption{
  Top 10 economies in the unweighted OTNs. According to Eq.~(\ref{Eq:node:criticality}), we calculate the criticality of each economy in the unweighted trade networks from 1988 to 2017 and arrange the index $C_i^{A}$ in descending order. The criticality rankings are shown every two years.}
      \resizebox{\textwidth}{40mm}{
    \begin{tabular}{ccccccc}
    \toprule
    Rank  & 1990  & 1993  & 1996  & 1999  & 2002  & 2005 \\
    \midrule
    1     & Japan & USA   & USA   & USA   & USA   & USA \\
    2     & Fmr Fed. Rep. of Germany & United Kingdom & Nigeria & Italy & China & Russian Federation \\
    3     & Malaysia & Mexico & United Kingdom & Venezuela & Australia & United Arab Emirates \\
    4     & Australia & Germany & France & Germany & Germany & France \\
    5     & Canada & Argentina & Germany & France & Italy & Germany \\
    6     & Singapore & Rep. of Korea & Netherlands & Albania & Kazakhstan & China \\
    7     & Rep. of Korea & China & Rep. of Korea & Russian Federation & France & Italy \\
    8     & Spain & Singapore & So. African Customs Union & Rep. of Korea & United Kingdom & United Kingdom \\
    9     & New Zealand & Malaysia & Algeria & Trinidad and Tobago & Belgium & Netherlands \\
    10    & Thailand & Australia & Austria & United Kingdom & Trinidad and Tobago & Singapore \\
    \midrule
    Rank  & 2008  & 2011  & 2014  & 2015  & 2016  & 2017 \\
    \midrule
    1     & USA   & USA   & USA   & Netherlands & Netherlands & Netherlands \\
    2     & Canada & Nigeria & China & USA   & USA   & USA \\
    3     & Russian Federation & China & Russian Federation & South Africa & Russian Federation & South Africa \\
    4     & China & Netherlands & South Africa & United Arab Emirates & South Africa & India \\
    5     & United Kingdom & United Kingdom & France & United Kingdom & China & China \\
    6     & France & South Africa & Netherlands & France & France & United Arab Emirates \\
    7     & India & India & Indonesia & Spain & Germany & United Kingdom \\
    8     & United Arab Emirates & Singapore & Singapore & China & India & Russian Federation \\
    9     & South Africa & Canada & Italy & Russian Federation & United Arab Emirates & France \\
    10    & Italy & Sweden & India & India & United Kingdom & Indonesia \\
    \bottomrule
    \end{tabular}}%
  \label{tab:node:criticalityRank10}%
\end{table*}%

\begin{table*}[!t]
  \centering
  \caption{
  Top 10 economies in the weighted OTNs. According to Eq.~(\ref{Eq:node:criticality}), we calculate the criticality of each economy in the weighted trade networks from 1988 to 2017 and arrange the index $C_i^{W}$ in descending order.}
        \resizebox{\textwidth}{40mm}{
    \begin{tabular}{ccccccc}
    \toprule
    Rank  & 1990  & 1993  & 1996  & 1999  & 2002  & 2005 \\
    \midrule
    1     & Japan & USA   & USA   & USA   & USA   & USA \\
    2     & Saudi Arabia & Saudi Arabia & Saudi Arabia & Saudi Arabia & Saudi Arabia & Saudi Arabia \\
    3     & Fmr Fed. Rep. of Germany & Japan & United Kingdom & Japan & Russian Federation & Russian Federation \\
    4     & United Arab Emirates & United Kingdom & Japan & Norway & Japan & China \\
    5     & Spain & Germany & Netherlands & Netherlands & United Kingdom & Japan \\
    6     & Iran  & Norway & Norway & United Kingdom & Norway & Brazil \\
    7     & Singapore & United Arab Emirates & Canada & Russian Federation & Rep. of Korea & Canada \\
    8     & Rep. of Korea & Canada & Russian Federation & Rep. of Korea & Netherlands & United Kingdom \\
    9     & United Kingdom & Rep. of Korea & Germany & Canada & China & Netherlands \\
    10    & Canada & Iran  & Nigeria & Germany & Indonesia & Norway \\
    \midrule
    Rank  & 2008  & 2011  & 2014  & 2015  & 2016  & 2017 \\
    \midrule
    1     & USA   & USA   & USA   & USA   & USA   & USA \\
    2     & Saudi Arabia & Saudi Arabia & Saudi Arabia & China & China & China \\
    3     & Russian Federation & China & Canada & Canada & Saudi Arabia & Saudi Arabia \\
    4     & Japan & Russian Federation & China & Saudi Arabia & Canada & Canada \\
    5     & China & Canada & Russian Federation & Russian Federation & Russian Federation & United Kingdom \\
    6     & Canada & Japan & United Kingdom & India & United Kingdom & Russian Federation \\
    7     & United Kingdom & United Kingdom & India & United Kingdom & Netherlands & India \\
    8     & Rep. of Korea & India & Netherlands & Netherlands & India & Netherlands \\
    9     & Netherlands & Rep. of Korea & Japan & Rep. of Korea & Rep. of Korea & Rep. of Korea \\
    10    & United Arab Emirates & Nigeria & Rep. of Korea & Iraq  & Japan & Iraq \\
    \bottomrule
    \end{tabular}}%
  \label{tab:node:weight:criticalityRank10}%
\end{table*}%

\subsubsection{Evolving criticality of each continent}

For the global OTNs, the criticality of each economy has essential reference value for the development of the economy's subsequent development strategy. We categorize economies in the OTNs by continents and calculate the criticality of each continent. The classification of economies by continents is a method that takes into account geographical, economic and political factors. This method is consistent with objective reality.

\begin{figure}[!t]
\centering
\includegraphics[width=0.95\linewidth]{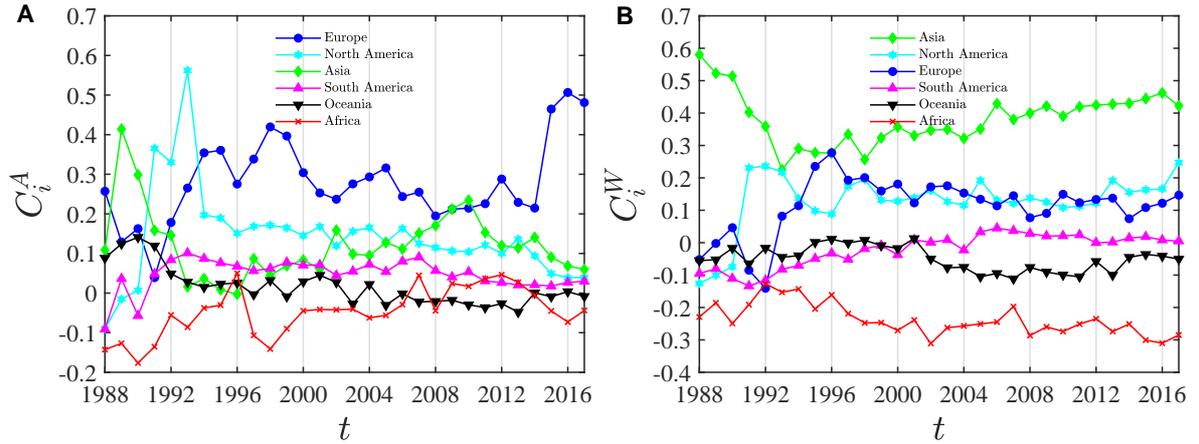}
  \caption{
  Criticality evolution of each continent in $A_t$ and $W_t$. The economies involved from 1988 to 2017 in OTNs are divided into six categories: Europe, Asia, Africa, Oceania, South America, and North America. The criticality values of the economies included in each continent are accumulated in each year, and the sum is used as the criticality value of each continent in that year. (A) and (B) are the evolution of the criticality of the continents in $A_t$ and $W_t$. Each different line represents the evolution of different continents. The legend is determined by the descending order of the cumulative criticality of the economies in each continent for 30 years.}
    \label{Fig:oil:nodes:criticality:continents}
\end{figure}

In Fig.~\ref{Fig:oil:nodes:criticality:continents}~(A), Europe ranks first in cumulative criticality. It can also be seen from the dark blue line that although the criticality of European economies has fluctuated since 1994, it has basically been above other continents, followed by North America and Asia. For weighted network criticality in Fig.~\ref{Fig:oil:nodes:criticality:continents}~(B), the top continent is Asia, followed by North America and Europe. Regardless of the type of networks, Africa ranks last in criticality. Such differences have significant relations with geographical location and political and economic development. In recent years, the global oil trade has undergone essential changes brought about by factors such as the increase in production as a result of the North American shale oil and gas revolution in 2014 and the economic sanctions imposed on Russia by Europe and the United States after the Ukraine crisis broke out.

\subsection{Evolving criticality of trade relationships}

\begin{table*}[!t]
  \centering
  \caption{
  Top 10 trade relationships in unweighted OTNs from 2008 to 2017. According to Eq.~(\ref{Eq:edge:criticality}), we calculate the criticality of each trading relationship in $A_t$, and arrange the index $C_{ij}^{A}$ in descending order. The criticality rankings are shown from 2008 to 2012.}
        \resizebox{\textwidth}{30mm}{
    \begin{tabular}{cccccc}
    \toprule
    Rank  & 2008  & 2009  & 2010  & 2011  & 2012 \\
    \midrule
    1     & Kyrgyzstan$\to$Kazakhstan & Finland$\to$Rep. of Korea & Guatemala$\to$USA & USA$\to$Venezuela & Croatia$\to$Netherlands \\
    2     & Romania$\to$Canada & Namibia$\to$Angola & Niger$\to$China & United Rep. of Tanzania$\to$Sweden & Serbia$\to$Croatia \\
    3     & Nicaragua$\to$Colombia & Niger$\to$China & Kenya$\to$Uganda & Guatemala$\to$USA & China$\to$Mali \\
    4     & Nigeria$\to$Senegal & Jordan$\to$Syria & Bangladesh$\to$Philippines & Saint Lucia$\to$United Kingdom & USA$\to$Venezuela \\
    5     & Guatemala$\to$USA & USA$\to$Jordan & Hungary$\to$Slovakia & Canada$\to$Russian Federation & Namibia$\to$South Africa \\
    6     & Ireland$\to$United Kingdom & Sweden$\to$Finland & Bahamas$\to$USA & Romania$\to$Azerbaijan & Russian Federation$\to$Kazakhstan \\
    7     & USA$\to$Libya & Ghana$\to$Spain & China$\to$Albania & Bouvet Island$\to$Nigeria & Netherlands$\to$Georgia \\
    8     & USA$\to$Grenada & Bangladesh$\to$Philippines & China$\to$Madagascar & USA$\to$Estonia & Romania$\to$Serbia \\
    9     & USA$\to$Palau & Andorra$\to$Nigeria & China$\to$Saint Kitts and Nevis & Bermuda$\to$Nigeria & Rep. of Korea$\to$Mauritius \\
    10    & USA$\to$Saint Kitts and Nevis & India$\to$Bhutan & China$\to$Burkina Faso & USA$\to$Bahamas & Cuba$\to$Netherlands \\
    \midrule
    Rank  & 2013  & 2014  & 2015  & 2016  & 2017 \\
    \midrule
    1     & Canada$\to$Russian Federation & Bosnia Herzegovina$\to$Indonesia & Netherlands$\to$USA & Netherlands$\to$USA & Netherlands$\to$USA \\
    2     & Ghana$\to$Togo & Eswatini$\to$South Africa & Spain$\to$Montenegro & France$\to$Cabo Verde & USA$\to$Guyana \\
    3     & Lithuania$\to$Poland & Bermuda$\to$USA & France$\to$Benin & USA$\to$Saint Kitts and Nevis & India$\to$Nepal \\
    4     & Poland$\to$Germany & Suriname$\to$USA & France$\to$Senegal & India$\to$Nepal & USA$\to$Saint Kitts and Nevis \\
    5     & Belize$\to$USA & USA$\to$Saint Kitts and Nevis & USA$\to$Saint Kitts and Nevis & USA$\to$Barbados & USA$\to$Argentina \\
    6     & Greece$\to$United Arab Emirates & USA$\to$Togo & USA$\to$Belize & USA$\to$Antigua and Barbuda & Spain$\to$Egypt \\
    7     & Netherlands$\to$Georgia & USA$\to$Montserrat & USA$\to$Iceland & USA$\to$North Macedonia & France$\to$Cote d'Ivoire \\
    8     & Netherlands$\to$Senegal & China$\to$State of Palestine & USA$\to$Tunisia & USA$\to$Senegal & United Arab Emirates$\to$Seychelles \\
    9     & Pakistan$\to$United Arab Emirates & China$\to$Guinea & United Kingdom$\to$Saint Vincent and the Grenadines & Germany$\to$Honduras & United Kingdom$\to$Uruguay \\
    10    & Uzbekistan$\to$Kazakhstan & Singapore$\to$Nepal & Rep. of Korea$\to$Madagascar & South Africa$\to$Eswatini & China$\to$Kenya \\
    \bottomrule
    \end{tabular}}%
  \label{tab:2008:2017edge:criticalityRank10}%
\end{table*}%

\begin{table*}[!t]
  \centering
  \caption{
  Top 10 trade relationships in weighted OTNs from 2008 to 2017. According to Eq.~(\ref{Eq:edge:criticality}), we calculate the criticality of each trading relationship in $W_t$ and arrange the index $C_{ij}^{W}$ in descending order. The criticality rankings are shown from 2008 to 2012.}
    \resizebox{\textwidth}{35mm}{
    \begin{tabular}{cccccc}
    \toprule
    Rank  & 2008  & 2009  & 2010  & 2011  & 2012 \\
    \midrule
    1     & Canada$\to$USA & Canada$\to$USA & Canada$\to$USA & Canada$\to$USA & Canada$\to$USA \\
    2     & Saudi Arabia$\to$USA & Netherlands$\to$Belgium & Netherlands$\to$Belgium & Netherlands$\to$Belgium & Netherlands$\to$Belgium \\
    3     & Netherlands$\to$Belgium & Venezuela$\to$USA & Saudi Arabia$\to$Japan & Saudi Arabia$\to$Japan & USA$\to$Canada \\
    4     & Venezuela$\to$USA & Saudi Arabia$\to$Japan & Mexico$\to$USA & Saudi Arabia$\to$USA & Saudi Arabia$\to$USA \\
    5     & Saudi Arabia$\to$Japan & Mexico$\to$USA & Venezuela$\to$USA & Mexico$\to$USA & Saudi Arabia$\to$Japan \\
    6     & United Arab Emirates$\to$Japan & USA$\to$Canada & Saudi Arabia$\to$USA & Venezuela$\to$USA & Saudi Arabia$\to$China \\
    7     & Mexico$\to$USA & Saudi Arabia$\to$USA & Nigeria$\to$USA & Saudi Arabia$\to$China & Mexico$\to$USA \\
    8     & Nigeria$\to$USA & Saudi Arabia$\to$China & Norway$\to$United Kingdom & Norway$\to$United Kingdom & Venezuela$\to$USA \\
    9     & Saudi Arabia$\to$Rep. of Korea & United Arab Emirates$\to$Japan & Saudi Arabia$\to$China & United Arab Emirates$\to$Japan & Saudi Arabia$\to$Rep. of Korea \\
    10    & USA$\to$Canada & Nigeria$\to$USA & Angola$\to$China & USA$\to$Canada & Angola$\to$China \\
    \midrule
    Rank  & 2013  & 2014  & 2015  & 2016  & 2017 \\
    \midrule
    1     & USA$\to$Canada & USA$\to$Canada & USA$\to$Canada & USA$\to$Canada & Canada$\to$USA \\
    2     & Canada$\to$USA & Canada$\to$USA & Canada$\to$USA & Canada$\to$USA & USA$\to$United Kingdom \\
    3     & Canada$\to$United Kingdom & Netherlands$\to$Belgium & Netherlands$\to$Belgium & Netherlands$\to$Belgium & USA$\to$Canada \\
    4     & Netherlands$\to$Belgium & Saudi Arabia$\to$USA & Saudi Arabia$\to$USA & Saudi Arabia$\to$USA & Netherlands$\to$Belgium \\
    5     & Saudi Arabia$\to$USA & Saudi Arabia$\to$Japan & Saudi Arabia$\to$China & Saudi Arabia$\to$Japan & Saudi Arabia$\to$Japan \\
    6     & Saudi Arabia$\to$Japan & Canada$\to$United Kingdom & Venezuela$\to$USA & Russian Federation$\to$China & Russian Federation$\to$China \\
    7     & Saudi Arabia$\to$China & Mexico$\to$USA & Mexico$\to$USA & Venezuela$\to$USA & Mexico$\to$USA \\
    8     & Mexico$\to$USA & Venezuela$\to$USA & Canada$\to$United Kingdom & Saudi Arabia$\to$China & Venezuela$\to$USA \\
    9     & Venezuela$\to$USA & Saudi Arabia$\to$China & United Arab Emirates$\to$Japan & Angola$\to$China & Saudi Arabia$\to$USA \\
    10    & United Arab Emirates$\to$Japan & United Arab Emirates$\to$Japan & Russian Federation$\to$China & Saudi Arabia$\to$Rep. of Korea & Indonesia$\to$USA \\
    \bottomrule
    \end{tabular}}%
  \label{tab:2008:2017edge:weight:criticalityRank10}%
\end{table*}%

\begin{figure}[!t]
\centering
\includegraphics[width=0.95\linewidth]{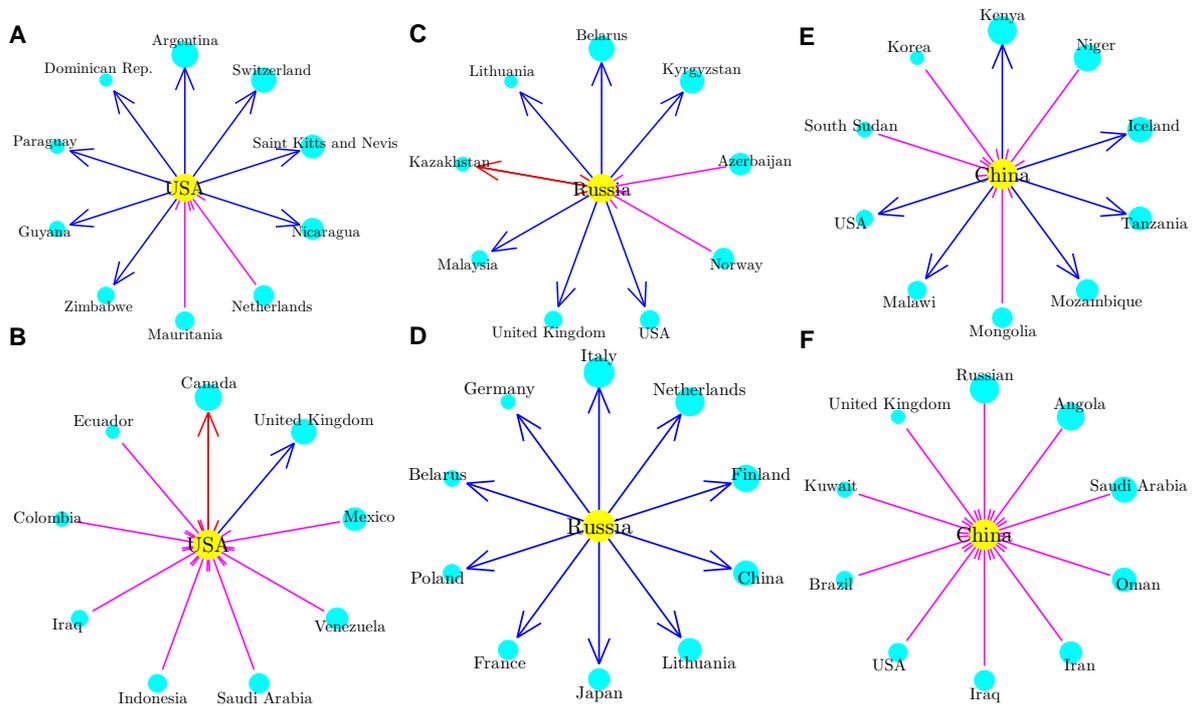}
  \caption{
  Critical trade relationships of USA, Russia and China in 2017. The first row (A,C,E) and the second row (B,D,F) are the top 10 critical relationships between the USA, Russian, and China in the unweighted and weighted trade networks. The central nodes represent the USA, Russia, and China, respectively. The nodes at the edges are the ten most criticality trade relationships. Import and export trades between countries are also recognized as the top 10 relationships at the same time, and the directions of arrows represent the directions of oil trade flows.  Bilateral trades are indicated by bidirectional arrows. The size of the nodes is based on the criticality of trade relationships, and they are arranged counterclockwise from small to large  the smaller the node, the lower the ranking.}
    \label{Fig:oil:edges:cri:2017}
\end{figure}

The rapid changes in the global oil trade are often reflected in the changing trade relationships of various economies. It is of practical significance to measure which trade relationships are more critical to the efficiency of trade networks. Based on network efficiency indexes, critical economies in the OTNs can be identified. According to Eq.~(\ref{Eq:edge:criticality}), we calculate the criticality of trade relationships in the networks. We simulate the impact on the efficiency of the trade networks after cutting off the trade relationship between two economies (such as economic sanctions in the real world), which provides a new method for measuring the criticality of oil trade relationships. Table~\ref{tab:2008:2017edge:criticalityRank10} and Table~\ref{tab:2008:2017edge:weight:criticalityRank10} show the top 10 critical trade relationships in the unweighted and weighted global OTNs from 2008 to 2017.

  It is found that there is a big difference in the recognition results between the unweighted and weighted trade networks. In the unweighted OTNs, the critical trade relationships are identified based on the network topology. From Table~\ref{tab:2008:2017edge:criticalityRank10}, the top 10 critical trade relationships in the past 10 years are very different. This illustrates the time-varying characteristics of the network structure. From 2015 to 2017, the most critical trade relationship is from the Netherlands to the USA in the unweighted trade networks.

  Due to the network weights, the critical relationships have changed less each year and many relationships have appeared in the last ten years, reflecting certain stability. For example, the trade relationship between Canada and the USA  ranks among the top 3 in almost all years. Relationships with the USA are almost over 50\% in the top 10 rankings. The remaining relationships also involve major oil import and export economies. The impact of weight on criticality is very significant.

  The focus of relationship criticality identification between the unweighted and weighted OTNs is different. The criticality of relationships in unweighted networks helps to eliminate the impact of weights and measure the structural criticality of the relationships. However, in the weighted networks, it helps to examine criticality based on the strength of the trade relationships. This is more practical and can provide decision support for the development of global OTNs.

Among the critical trade relationships, we select the top 10 trade relationships in 2017 related to the USA, Russia and China. In Fig.~\ref{Fig:oil:edges:cri:2017}, we find that the critical relationships of the USA, Russia and China identified in network $A_t$ and $W_t$ are very different. The red arrows in the figure indicate that the bidirectional trade relationships between two economies are critical, such as between the USA and Canada, and between Russia and Kazakhstan.

In 2017, the USA shook off its dependence on traditional oil exporters and built new trade partners. While promoting energy self-sufficiency, the USA has established stable and long-term trade relationships with Canada, Mexico and other American economies. These economies supply high-quality and cheap crude oil and become the self-sufficient back garden of the USA. Among the relationships, Canada is the largest trade partner of the USA. They have maintained a stable and close trade-friendly relationship.

Dependence between Russia and its trading partners had been low due to falling oil prices and economic sanctions, but it is now recovering. As one of the largest oil exporters in the world, Russia has continuously adapted to the global environment through reforms and adjustments, and the social and economic trends that had worsened due to continual sanctions and low oil prices have been contained. The import decline in the USA leaves room for import trade for China, India and other oil-consuming economies. China has gradually become the largest oil importer in the world. Due to changes in the oil supply structure, import economies need to develop cooperative relationships with emerging North American exporters, and strengthen relationships with economies such as Nigeria and Angola, carry out variable diplomatic cooperation and further expand sources of importing.

\subsection{Robustness evolution of the global OTNs}

\subsubsection{Robustness under economy attack strategies}

The robustness of the OTN means that the network can guarantee certain structural integrity and functional capability in the event of economy failure (national bankruptcy) and trade relationship failure (economic sanctions). Network robustness is an essential dynamic characteristic of systems. Robust analysis of the OTNs can help understand the mechanism and rules of the failure or collapse of the oil trade system. Furthermore, it can help find better ways to prevent risks and maintain the stability and healthy development of the networks.

Most of the economies in the OTNs have a relatively small degree, which means that most economies have trade relationships with only a few economies. Other economies, such as the USA and Russia, have a much larger number of trade relationships. These economies can be called hub economies. Such a network structure results in higher network robustness when economies are randomly removed. If a deliberate attack (based on specific deterministic indicators) is adopted, the OTNs will be more vulnerable.

\begin{figure}[!t]
\centering
\includegraphics[width=0.95\linewidth]{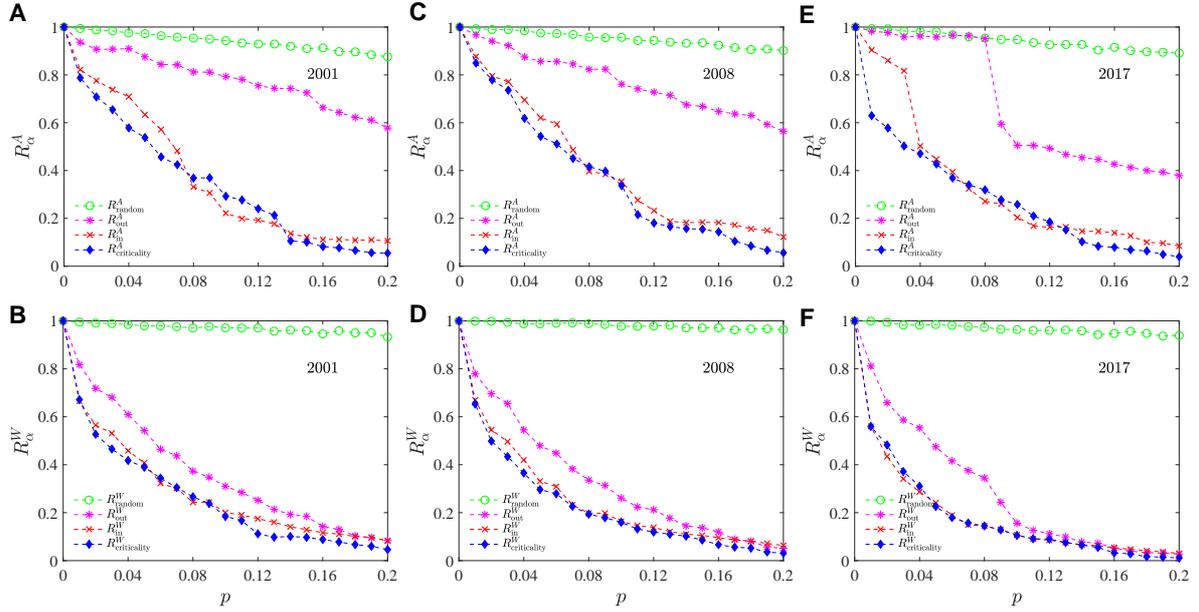}
  \caption{
  Robustness of the OTNs under economy attacks in 2001, 2008 and 2017. The figure shows the robustness evolution under the four economies attack strategies. The four attack strategies are as follows. (1) criticality: removing economies from the network with the proportion of $p$ in the criticality rankings; (2) random: deleting the corresponding number of nodes randomly; (3) in: removing economies from the network with the proportion of $p$ in the import volume rankings; (4) out: removing economies from the network with the proportion of $p$ in the import volume rankings. The three graphs in the first row (A), (C) and (E) from left to right are the robustness evolution in 2001, 2008, and 2017. The second row (B), (D) and (F) are the robustness evolution of the weighted networks.}
    \label{Fig:oil:nodes:robustness20010817}
\end{figure}

In order to study the robustness of the OTN $A_t$ and $W_t$, according to Eq.~(\ref{Eq:robustness:node:edge:attack}), we calculate the robustness under different attack strategies and attack levels. The result is shown in Fig.~\ref{Fig:oil:nodes:robustness20010817} and the findings can be summarized as follows.

  In unweighted or weighted OTNs, the robustness under random attack and the robustness under other deliberate attack strategies are quite different. The decrease of the robustness after the random deletion of the economies is almost linear with the increase of the attack degree $p$, and the decrease of the robustness is small. For unweighted trade networks, the random deletion of 20\% of economies reduces the robustness by about 10\%. However, for weighted networks, the decrease of robustness is smaller, about 5\%. When economies are deliberately attacked, the robustness of the trade network decreases sharply as the attack intensifies. When the attack strategies based on the export volumes of economies occurs, the decrease in robustness is much smaller than that from the attack strategies based on the import volumes and criticality ranking of economies. The scale-free nature of the OTNs \citep{Barabasi-Albert-1999-Science} determines the different trends of robustness changes under random attacks and different deliberate attacks.

  In Fig.~\ref{Fig:oil:nodes:robustness20010817}~(A,C,E), the level of network robustness changes is relatively similar as the attack intensifies for the unweighted network in 2001 and 2008. Under the random attack, the decrease of the robustness level is the smallest, and the robustness under the export volumes attack is relatively gentle. Under the import volumes attack and the criticality ranking attack, the robustness drops significantly. In 2017, robustness is different from the gentle decline in previous years. Under the attack of import and export volumes, when the attack level reached a certain level, the robustness would be significantly reduced. This change is related to structural changes that may be related to unweighted OTNs. Due to the influence of the weights of trade relationships in Fig.~\ref{Fig:oil:nodes:robustness20010817}~(B,D,F),  the trends of robustness in the three years are similar for the weighted network $W_t$.

The deliberate attacks have a more significant impact on the robustness than the random attack. The robustness will decline significantly as the number of attacked economies increase. The reason is that economies with higher criticality or import and export trade volumes have more trade relationships and total trade value. These economies are essential bridges for the flow of OTNs. The more attacks on these economies, the more likely they are to be cut off from some other small economies, thus even isolating those small economies and leading toa sharp decline in network robustness. In Fig.~\ref{Fig:oil:nodes:robustness20010817}, the network robustness value is higher under random economy attack.

\subsubsection{Robustness under failure of trade relationships}

Today's pattern of OTNs is still in the process of change. Both the attack of the economy and relationship constitute important research issues. The political status and trade policies of various economies are always in the process of change. Adjustments of trade policies between economies are likely to cause the invalidation of trade relationships. Therefore, the robustness research under the attacks of the trade relationships in the OTNs is meaningful.

\begin{figure}[!t]
\centering
\includegraphics[width=0.95\linewidth]{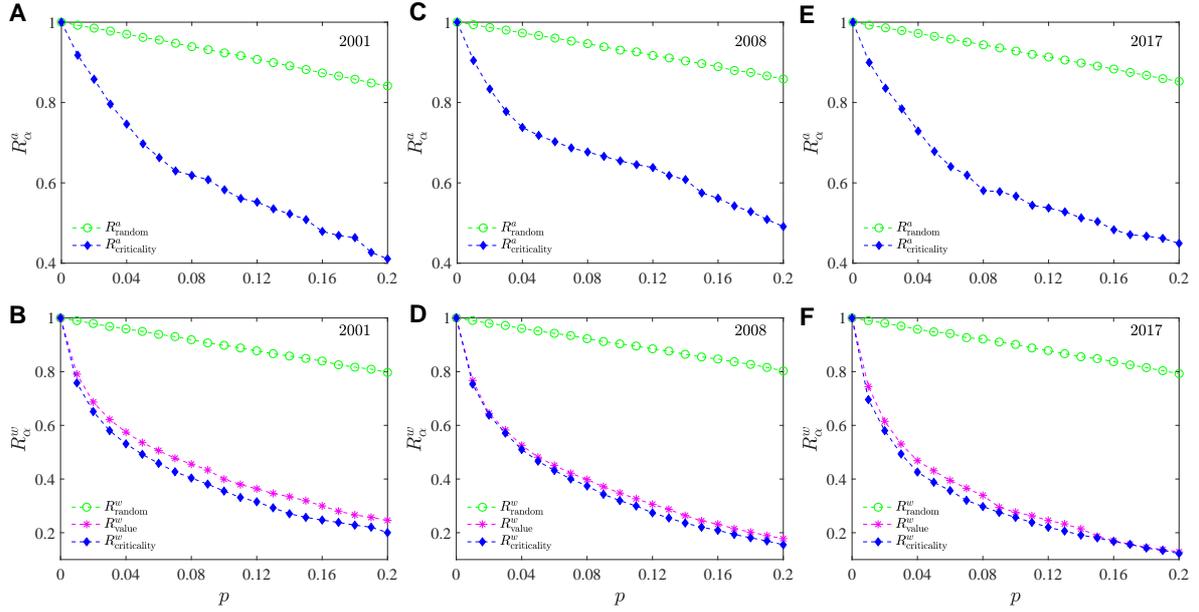}
  \caption{
  Robustness of the OTNs in the failure of trade relationships in 2001, 2008 and 2017. (A), (C), and (E) show the changes in the robustness of the unweighted OTNs in different years under different attack strategies of trade relationship failure. The two failure strategies are as follows: (1) criticality: deleting the trade relationships in descending order of the criticality of trade relationships. (2) random: deleting the corresponding number of trade relationships randomly. (B), (D) and (F) show the evolution of the robustness of the weighted OTNs. The first two attack strategies are the same as those for unweighted networks. The $R_{\rm{value}}$ indicates that the trade relationships are deleted in descending order of weights. The horizontal axis $p$ is the proportion of the deleted trade relationships, and the vertical axis is the robustness $R$. The upper right corner indicates the year of the oil network.}
    \label{Fig:oil:edges:robustness20010817}
\end{figure}

According to Eq.~(\ref{Eq:robustness:node:edge:attack}), we calculate the robustness of relationships under different trade relationships attack strategies and attack levels. Fig.~\ref{Fig:oil:edges:robustness20010817} shows the robust results of the OTNs in 2001, 2008 and 2017.

From Fig.~\ref{Fig:oil:edges:robustness20010817}~(A,C,E) and Fig.~\ref{Fig:oil:edges:robustness20010817}~(B,D,F), it can be seen that for the unweighted trade networks, we only adopt two strategies: random attack and attack based on criticality rankings. For weighted networks, we also consider attack based on the weight of relationship rankings. The robustness of the OTNs under random attack is very similar, and its value is also high. Nevertheless, for unweighted networks, compared with the economy attack in Fig.~\ref{Fig:oil:nodes:robustness20010817}, the decrease in robustness under relationship attacks is slight. The robustness drops to 0.4 after removing 20\% of trade relationships. For the weighted network in Fig.~\ref{Fig:oil:edges:robustness20010817}~(B,D,F), the robustness under the weight attack strategy and the robustness under the criticality strategy are very close, but the robustness decline is faster under the criticality attack.

\section{Discussion and summary}
\label{S1:Conclude}

In this study, we construct OTNs for each year based on the UN Comtrade data from 1988 to 2017 and introduce network efficiency indexes for measuring the efficiency of trade flows in the global oil trade market. We construct two types of OTNs: unweighted and weighted. We define not only the criticality index of each economy and trade relationship, but also the robustness indexes of the networks under the economy and trade relationship attack.

According to the results listed in the empirical analysis section, we make the following management recommendations. Firstly, more attention should be paid to the critical economies and trade relationships identified in the OTNs, because they will significantly affect the function of the OTN, such economies as the USA, China and Saudi Arabia and the relationships of them for instance. Secondly, in order to improve the robustness of the OTNs, economies need to establish more trade relationships, especially the relationship of higher criticality. Thirdly, the economies that have more relationships and large volumes can increase network efficiency and robustness under random attacks. However, at the same time, such properties will increase the risk of a sudden reduction of robustness when deliberate attacks occur. Once such an economy has a problem, there will a cascading effect affecting other economies and even the development of global oil trade. Since today's oil trade situation is changing rapidly, governments and policymakers need to be vigilant, making efforts to reduce the impact of sudden changes on the economy and trade, and seek a balance between network efficiency and robustness.

There are still many research directions of value and potential. Firstly, acquiring more detailed data, for example, the political and economic indicators of various economies in the OTN, which can be combined with the value of oil trade, will provide a new reference for measuring the weight of the trade relationships and then can be used to measure the efficiency and robustness in the weighted network. Secondly, the efficiency indicators of OTNs proposed here can be applied to other trading systems, such as gas, coal and other commodity trading systems, to identify critical economies and trade relationships. Thirdly, the balance between robustness and network efficiency in each system awaits to be found; an analysis of different systems will provide theoretical support for the development of each system.

\section*{Acknowledgment}

This work was supported by the National Natural Science Foundation of China under grants U1811462, the Shanghai Outstanding Academic Leaders Plan, and the Fundamental Research Funds for the Central Universities.

%

\end{document}